\documentclass{aastex}
\usepackage{emulateapj5}

\newcommand{\ba}{\begin{eqnarray}}
\newcommand{\ea}{\end{eqnarray}}
\newcommand{\be}{\begin{equation}}
\newcommand{\ee}{\end{equation}}
\newcommand{\bfv}{{\bf v}}
\newcommand{\bu}{{\bf u}}

\newcommand{\tomega}{{\tilde\omega}}
\newcommand{\rmRe}{{\rm Re}}

\def\bomega{{\mbox{\boldmath $\omega$}}}
\def\bmu{{\mbox{\boldmath $\mu$}}}
\def\cD{{\cal D}}
\def\cF{{\cal F}}
\def\go{\mathrel{\raise.3ex\hbox{$>$}\mkern-14mu
             \lower0.6ex\hbox{$\sim$}}}
\def\lo{\mathrel{\raise.3ex\hbox{$<$}\mkern-14mu
             \lower0.6ex\hbox{$\sim$}}}

\newenvironment{inlinefigure}{
\medskip
\def\@captype{figure}
\noindent\begin{minipage}{0.999\linewidth}\begin{center}}
{\end{center}\end{minipage}\medskip}

\begin{document}

\title{Wave Excitation in Disks Around Rotating Magnetic Stars}
\author{Dong Lai\altaffilmark{1} and Hang Zhang\altaffilmark{2}}
\altaffiltext{1}{Department of Astronomy,
Cornell University, Ithaca, NY 14853. Email: dong@astro.cornell.edu}
\altaffiltext{2}{Department of Physics, Nanjing Normal University, Nanjing, JS 210097, 
P.R.~China}

\begin{abstract}
The accretion disk around a rotating magnetic star (neutron star, 
white dwarf or T Tauri star) is subjected to periodic
vertical magnetic forces from the star, with the forcing frequency equal to 
the stellar spin frequency or twice the spin frequency. This gives rise
bending waves in the disk that may influence the variabilities of the system.
We study the excitation, propagation and dissipation of these waves
using a 
hydrodynamical model coupled with a generic model description of
the magnetic forces. The $m=1$ bending waves are excited at the 
Lindblad/vertical resonance, and propagate either to larger radii or 
inward toward the corotation resonance where dissipation takes place.
While the resonant torque is negligible compared to the accretion torque,
the wave nevertheless may reach appreciable amplitude and can cause or
modulate flux variabilities from the system. We discuss 
applications
of our result to the observed quasi-periodic oscillations from 
various systems, in particular neutron star low-mass X-ray binaries.
\end{abstract}

\section{Introduction}

Disk accretion onto magnetic central objects occurs in a variety of
astrophysical systems, ranging from classical T Tauri stars (e.g.,
Bouvier et al.~2007a), cataclysmic variables (intermediate polars;
e.g. Warner 2004), to accretion-powered X-ray pulsars (e.g. Lewin \&
van der Klis 2006). The basic picture of the disk-magnetosphere
interaction is well known: The stellar magnetic field disrupts the
accretion flow at the magnetospheric boundary and funnels the plasma
onto the polar caps of the star or ejects it to infinity.
The magnetosphere boundary is located where the magnetic and plasma
stresses balance, 
\be
r_m =\xi\,[\mu^4/(GM\dot M^2)]^{1/7},
\ee
where $M$ and $\mu$ are the mass and magnetic moment of the central
object, $\dot M$ is the mass accretion rate 
and $\xi$ is a dimensionless constant of order 0.5-1.
The funnel flow occurs when $r_m$ is less 
than the corotation
radius $r_c$ (where the disk rotates at the same rate as the
star). For $r_m\go r_c$, centrifugal forces may lead to ejection of
the accreting matter (``propeller'' effect).
Over the last several decades, numerous theoretical studies
have been devoted to understand the interaction between accretion
disk and magnetized star (e.g., Ghosh \& Lamb 1979;
Aly 1980;
Lipunov \& Shakura 1980; 
Anzer \& B\"orner 1983; Arons 1993;
Aly \& Kuijpers 1990; Spruit \& Taam 1993;
Shu et al.~1994,2000;
Wang 1995; 
Lovelace et al.~1995,1999; Li et al.~1996;
Campbell 1997; Lai 1998,1999; 
Terquem \& Papaloizou 2000; Shirakawa \& Lai 2002a,b; Pfeiffer \& Lai 2004;
Uzdensky 2004; Matt \& Pudritz 2005).  There have also been many
numerical simulations on accretion onto magnetic stars, 
with increasing sophistication and realism
(e.g., Stone \& Norman 1994; Hayashi et al.~1996; Goodson et al.~1997;
Miller \& Stone 1997; 
Fendt 2003; Romanova et
al.~2003,2006; Ustyugova et al.~2006; Long et al.~2007).

In this paper we carry out an analytical study of wave excitation 
in disks around rotating magnetic stars. The idea is very simple:
A rotating dipole, inclined relative to the stellar spin axis, 
induces a time-dependent magnetic force on the disk. 
The forcing frequency normally equals to the spin frequency $\omega_s$, 
but we show that under certain conditions the magnetic force could
also have a $2\omega_s$ component (see \S 2). The magnetic forces
excite $m=1$ density/bending waves in the disk, most prominently at the
Lindblad/vertical resonance. This magnetically driven resonance
was noted before (Lai 1999) but not studied.
Since details of the magnetic field -- disk interaction are
complex (e.g., the interaction depends on magnetic field 
dissipation/reconnection), we will
model the magnetic force using simple models that,
we conjecture,
capture the essential dynamical feature of the system. 
Petri (2005) examined several aspects of the forced oscillations in
magnetized disks, but did not study the resonantly excited bending waves
and their evolution that are of interest in this paper
(see also related work by Agapitou et al.~1997; Terquem \& Papaloizou 2000).

Wave excitations in magnetic disks may be relevant to understanding
the flux variabilities that have been observed in many accreting systems,
from pre-main-sequence stars (e.g., O'Sullivan er al.~2005, 
Bouvier et al.~2007b and references therein) to accreting magnetic 
white dwarfs and neutron stars.
It has been suggested that the phenomenology of the 
qausi-periodic brightness modulations in CVs might be explained by 
magnetically excited traveling waves in the disk (Warner 2004).
Of great interest are the high-frequency (kHz) quasi-periodic
oscillations (QPOs) observed in the X-ray fluxes of more than 20
neutron stars in Low-Mass X-ray Binaries (LMXBs) (see van der Klis 2006
for a review). These kHz QPOs often occur in pairs, with intriguing
correlation between the upper-frequency $\nu_u$, the 
lower-frequency $\nu_l$ and the spin frequency $\nu_s$. 
In particular, in the accreting millisecond pulsar SAX J1808.4-3658,
$\nu_u=694\pm 4$~Hz, $\nu_l=499\pm 4$~Hz, and $\nu_u-\nu_l$ equals
$\nu_s=401$~Hz to within a few Hz (Wijnands et al.~2003).
In another accreting millisecond pulsar XTE J1807.4-294, 
$\nu_u-\nu_l$ remains constant with an average value of  
$205\pm 6$~Hz, approximately equal to $\nu_s=191$~Hz, even
when both $\nu_l$ and $\nu_u$ vary over a range of more than 
$200$~Hz (Linares et al.~2005). In some systems, however,
$\nu_u-\nu_l$ are found to vary significantly (e.g., 
Circinus X-1, Boutloukos et al.~2006). In general, there 
is some evidence that $(\nu_u-\nu_l)/\nu_s$ clusters around 0.5 and
1 in those systems (about 10) for which these quantities have been measured,
indicating that the neutron star spin plays an important
role, although this issue is still under debate (see
Yin et al.~2007; Mendez \& Belloni 2007). It is also of interest to 
note that recent observations 
(Casella et al.~2007; Altamirano et al.~2007) 
suggest that many neutron stars in LMXBs have
non-negligible (but complex) magnetic fields. 

Our paper is organized as follows. In \S 2 we discuss
the generic forms of magnetic forces on a disk around a rotating
magnetic dipole. The basic equations governing the hydrodynamical 
response of a disk to an external periodic force are presented in \S 3.
\S 4 focuses on wave dynamics around the Lindblad/vertical resonance
and corotation resonance and \S 5 gives the global solution of the
disk response to the periodic magnetic force. Some 
applications of our result to QPOs in LMXBs 
are discussed in \S 6.

\section{Periodic Magnetic Forces on the Disk}


The interaction between the dipole magnetic field of a star and
its accretion disk is complex, and many papers have been
published on it (\S 1). 
To focus on the dynamics of waves driven by magnetic
forces, we consider two simple analytical models, representing two
limiting behaviors of the disk. The basic geometric setup is as
follows:
The disk lies in the $xy$-plane with its normal vector long
the $z$-axis; the spin ($\bomega_s$) 
of the star is inclined with respect to
the $z$-axis by an angle $\beta$ and lies in the $yz$-plane.
Because of the possibility of 
large-scale disk warping driven by the (static) stellar magnetic field, 
we will allow for $\beta\neq 0$; Lai 1999, Pfeiffer \& Lai 2004; 
see below). The stellar dipole $\bmu$ rotates around $\bomega_s$, 
and the angle between $\bmu$ and $\bomega_s$ is $\theta$, such that 
$\bmu$ varies in time according to
$\bmu=\mu\,\left(\sin\theta\cos\varphi_\mu\,\hat x
+\sin\theta\sin\varphi_\mu\,\hat y+\cos\theta\,\hat z\right).$

\subsection{Simple Model}

First consider the limiting case where the disk is a perfect conductor 
and has no large-scale 
magnetic field of its own. The inner radius of the disk is located at $r=r_m$.
The magnetic field produced by the stellar dipole cannot 
penetrate the disk, and a diamagnetic
surface current is induced. Aly (1980) has found the 
exact analytic solution for the magnetic field outside the thin disk.
At a point $(r,\phi,z=0)$ (cylindrical coordinates) on the disk 
surface ($z=0,~r>r_m$), the field is given by 
\ba
B_r&=&{2\mu\over r^3}\sin\chi\cos(\varphi-\varphi_\mu)\mp{4\mu\over\pi r^3
\cD}\cos\chi,\label{br1}\\
B_\varphi&=&{\mu\over r^3}\sin\chi\sin(\varphi-\varphi_\mu),\label{bphi1}
\ea
and $B_z=0$, where $\chi=\chi(t)$ is the angle between $\bmu$ and the z-axis.
In eq.~(\ref{br1}), the upper (lower) sign applies to the upper
(lower) disk surface, and 
$\cD={\rm max}\left(\sqrt{{r^2/r_m^2}-1},
\sqrt{{2h/r_m}}\right)$,
where $h\ll r_m$ is the half-thickness of the disk. 
The vertical magnetic force per unit area on the disk is 
\be
F_z={2\mu^2\over\pi^2r^6\cD}\sin 2\chi\,\cos(\varphi-\varphi_\mu).
\label{force1}
\ee
Using the relations between various angles,
we have
\be
F_z=\rmRe\sum_{\omega_f}F_{\omega_f}(r)\exp(i\phi-i\omega_f t),
\label{eq:force}\ee
where $\omega_f=0,\pm\omega_s,\pm 2\omega_s$, with 
\ba
F_0(r)&=&-i\cF_D\,(2-3\sin^2\theta)\,\sin 2\beta,\nonumber\\
F_{\omega_s}(r)&=&\cF_D\sin 2\theta \,(\cos\beta+\cos 2\beta),\nonumber\\
F_{-\omega_s}(r)&=& \cF_D \sin 2\theta\,(\cos\beta-\cos 2\beta),\nonumber\\
F_{2\omega_s}(r)&=& -i\cF_D\sin^2\theta\,\sin\beta\,(\cos\beta+1),\nonumber\\
F_{-2\omega_s}(r)&=& -i\cF_D\sin^2\theta\,\sin\beta\,(\cos\beta-1),
\ea
and
\be
\cF_D\equiv 
{\mu^2\over\pi^2r^6\cD}.
\ee


The disk is unlikey to be completely diamagnetic: various instabilities
and dissipation mechanisms will allow some of the stellar 
magnetic field to partially penetrate the disk. As a simple model,
we decompose the vertical stellar dipole field on the disk into 
a static component and a time-varying component:
\be
\! B_z^{(0)}=-{\mu\over r^3}\cos\chi=
{\mu\over r^3}\left(\sin\beta\sin\theta\sin\omega_s t
-\cos\beta\cos\theta\right).
\label{bz0}\ee
We assume that the static field penetrates the disk, while the
the variable field is shielded out of the disk by a screening current.
Because of the shear between the disk and the plasma outside the disk, 
the threaded vertical field is wound to produce an azimuthal field that 
has different signs on the upper and lower surfaces of the disk.
We thus adopt the following {\it ansatz} for the magnetic field in the disk
(see Lai 1999)
\ba
\!\!\!B_r&=&{2\mu\over r^3}\sin\chi\cos(\varphi\!-\!\varphi_\mu)\!\pm
\!{4\mu\over\pi
r^3\cD}\sin\beta\sin\theta\sin\omega_s t,\label{br3}\\
\!\!\!
B_\varphi&=&{\mu\over r^3}\sin\chi\sin(\varphi\!-\!\varphi_\mu)\!\pm\!
\zeta{\mu\over
r^3}\cos\beta\cos\theta,\label{bphi3}\\
\!\!\!B_z&=&-{\mu\over r^3}\cos\beta\cos\theta,\label{bz3}
\ea
where $\zeta$ is a positive constant of order unity.
The vertical magnetic force (per unit area) on the disk is given by
\ba
&& F_z=-{4\mu^2\over\pi^2 r^6\cD}\sin\beta\sin\theta
\sin\omega t\sin\chi\cos(\varphi-\varphi_\mu) \nonumber\\
&&\qquad\quad -{\zeta\mu^2\over 2\pi r^6}\cos\beta\cos\theta\sin\chi
\sin(\varphi-\varphi_\mu).
\label{force3}\ea
This force can be written in the form of eq.~(\ref{eq:force}), with
\ba
&&\!\!F_0(r)\!=\!i\cF_D\,\sin^2\theta\,\sin 2\beta+\cF_T\,\cos^2\theta\,
\sin 2\beta,\nonumber\\
&&\!\!F_{\omega_s}(r)\!=\! \cF_D\,\sin 2\theta\,\sin^2\beta+
i\cF_T\,\sin 2\theta\,\cos\beta\cos^2{\beta\over 2},\nonumber\\
&&\!\!F_{-\omega_s}(r)\!=\! -\cF_D \sin 2\theta\,\sin^2\beta+
i\cF_T\sin 2\theta\cos\beta\sin^2{\beta\over 2},\nonumber\\
&&\!\!F_{2\omega_s}(r)\!=\! -i\cF_D\sin^2\theta\,\sin\beta\,(2+\cos\beta),
\nonumber\\
&&\!\!F_{-2\omega_s}(r)\!=\! i\cF_D\sin^2\theta\,\sin\beta\,(2-\cos\beta),
\ea
where
\be
\cF_T\equiv {\zeta\mu^2\over 4\pi r^6}.
\ee

\subsection{General Property of Magnetic Forces}

The magnetic forces discussed 
above share some general
properties. The zero-frequency force components $F_0$ have been shown to
lead to secular warping and precession of the disk (Lai 1999; 
Shirakawa \& Lai 2002a,b; Pfeiffer \& Lai 2004): The force resulting
from threaded field ($\propto \cF_T$) induces a warping instability, while
the force resulting from the dielectric response of the disk ($\propto \cF_D$)
gives rise to precession. The negative-frequency components ($F_{-\omega_s}$
and $F_{-2\omega_s}$) average to zero on the timescale of spin period. 

The positive-frequency forces, $F_{\omega_s}$ and $F_{2\omega_s}$, are
of interest for this paper.  They are generally present for 
non-zero magnetic field inclination angle $\theta$. Note that 
the $2\omega_s$ component arises because the dipole field
varies as $\sin\omega_s t$ or $\cos\omega_s t$, and the induced 
screening current also varies as $\sin\omega_s t$ or $\cos\omega_s t$,
and $F_{2\omega_s}$ is present only when $\sin\beta\neq 0$.

We note that periodic magnetic forces in the radial direction are also 
present in some disk models, but we we will focus on the dynamics of 
the disk under the periodic vertical force in this paper.

\section{Basic Equations: Disk dynamics under a periodic non-potential
forcing}

The key approximation underlying our study is that we treat the
waves in the disk using pure hydrodynamics, with the magnetic effect
entering only as external forces. This simplification allows
us to focus on the dynamics of wave excitation, propagation and
dissipation.  As we will see, the wave dynamics is sufficiently
complex and subtle even without magnetic fields,
and we believe that such a 
hydrodynamical treatment can be considered a useful first step. 
Future study using
MHD will be of great interest (and obviously much more difficult).  
We conjecture 
that in a real system, the wave dynamics will not differ
qualitatively from that described in this paper since the the waves of
interest here are basically sound waves modified by disk rotation, and
in the presence of magnetic fields the sound waves simply become fast
magnetosonic waves (with the sound speed replaced by fast wave speed).

In this section we develop the general hydrodynamical equations
for three-dimensional (3D) disks under generic non-potential forces
(cf. Zhang \& Lai 2006). 
We consider a geometrically thin gas disk and adopt cylindrical
coordinates $(r,\varphi,z)$.  The unperturbed disk has velocity
$\bfv_0=(0,r\Omega,0)$, where the angular velocity $\Omega=\Omega(r)$
is taken to be a function of $r$ alone. The disk is assumed to be
isothermal in the vertical direction 
and non-self-gravitating. 
Thus the vertical density profile is
given by 
\be 
\rho_0(r,z)={\sigma\over\sqrt{2\pi}h}\exp (-Z^2/2),\quad
{\rm with}~~ Z=z/h 
\label{eq:rho0}\ee 
where $h=h(r)=c/\Omega_\perp$ is the disk scale
height, $c=c(r)=\sqrt{p_0/\rho_0}$ is the isothermal sound speed,
$\sigma=\sigma(r)=\int dz\,\rho_0$ is the surface density, and
$\Omega_\perp$ is the vertical oscillation frequency of the disk.

We now consider perturbation of the disk driven by an external force
(per unit mass) ${\bf f}$. For simplicity, we shall assume that the 
perturbation is isothermal, so that the (Eulerian) density and pressure 
perturbations are related by $\delta P=c^2\delta\rho$.
The linear perturbation equations read
\ba
&&{\partial \bu\over\partial t}+(\bfv_0\cdot\nabla)\bu+(\bu\cdot\nabla)\bfv_0
=-\nabla\eta+{\bf f},\label{eq:u}\\
&& {\partial\rho\over\partial
t}+\nabla\cdot(\rho_0\bu+\bfv_0\delta\rho)=0, 
\label{eq:rho}\ea 
where $\bu=\delta\bfv$ is the (Eulerian) density perturbation,
and $\eta\equiv\delta p/\rho_0$ is the enthalpy perturbation.
Without loss of generality, each perturbation variable and the external
force are assumed to depend on $t$ and $\varphi$ as
$\exp(im\varphi-i\omega t)$,
where $m>0$ is an integer and $\omega$ is allowed 
to be either positive or negative, corresponding to the
prograde or retrograde wave, respectively. 
Equations (\ref{eq:u}) and (\ref{eq:rho}) then reduce to
\ba
&&\!\!\!-i\tomega u_r-2\Omega u_\varphi=-{\partial\over\partial r}\eta+f_r,
\label{eq:fluid1}\\
&&\!\!\!
-i\tomega u_\varphi +{\kappa^2\over 2\Omega}u_r=-{im\over r}\eta+f_\varphi,
\label{eq:fluid22}\\
&&\!\!\!-i\tomega u_z=-{\partial \over\partial z}\eta+f_z,\label{eq:fluid3}\\
&&\!\!\!-i\tomega {\rho_0\over c^2}\eta+{1\over r}{\partial\over\partial
r} (r\rho_0 u_r)+{im\over r}\rho_0 u_\varphi +{\partial\over\partial
z}(\rho_0 u_z)=0. \label{eq:fluid}\ea 
Here $\tomega$ is the ``Doppler-shifted'' frequency
\be
\tomega=\omega-m\Omega,
\ee
and $\kappa$ is the epicyclic frequency, defined by 
\be \kappa^2={2\Omega\over r}{d\over
dr}(r^2\Omega). \ee 
In this paper we will consider cold, (Newtonian)
Keplerian disks, for which the three characteristic frequencies,
$\Omega,\Omega_\perp$ and $\kappa$, are identical and equal to the
Keplerian frequency $\Omega_K=(GM/r^3)^{1/2}$. 
However, we continue to use different notations ($\Omega,\Omega_\perp,
\kappa$) for them in our treatment below when possible so that 
the physical origins of various terms are clear.

To separate out the $z$-dependence, 
we expand the perturbations with Hermite polynomials $H_n$
(cf.~Okazaki et al.~1987; Kato 2001)
\ba
\left[\begin{array}{c}
f_r(r,z)\\
f_\varphi(r,z)\\
\eta(r,z)\\
u_r(r,z)\\
u_\varphi(r,z)\end{array}\right]
&=& \sum_n \left[\begin{array}{c}
f_{rn}(r)\\
f_{\varphi n}\\
\eta_n(r)\\
u_{rn}(r)\\
u_{\varphi n}(r)\end{array}\right] H_n(Z),\nonumber\\
\left[\begin{array}{c}
f_z(r,z)\\
u_z(r,z)\end{array}\right]
&=& \sum_n \left[\begin{array}{c}
f_{zn}(r)\\
u_{zn}(r)\end{array}\right] H_n'(Z),
\label{eq:expand}\ea 
where $H'_n=dH_n/dZ$, and the Hermite polynomial is defined by
$H_n(Z)\equiv(-1)^n e^{Z^2/2}d^n(e^{-Z^2/2})/dZ^n$.
Note that since $H_1=Z$, $H_2=Z^2-1$, the $n=1$ mode 
coincides with the bending mode studied by Papaloizou \& Lin (1995) 
(who considered disks with no resonance), and the $n=2$ mode is similar 
to the mode studied by Lubow (1981).
With the expansion in (\ref{eq:expand}), the fluid equations (\ref{eq:fluid})
become
\ba &&-i\tomega u_{rn}-2\Omega u_{\varphi n}=
-{d\over dr} \eta_n+{n\mu\over r}\eta_n\nonumber\\
&& \qquad \qquad \qquad +{(n+1)(n+2)\mu\over r}\eta_{n+2}+f_{rn},\label{u1}\\
&&-i\tomega u_{\varphi n} +{\kappa^2\over 2\Omega}u_{rn}=-{im\over r}\eta_n
+f_{\varphi n},\label{u2}\label{eq:fluida}\\
&&-i\tomega u_{zn}=-{\eta_n\over h}+f_{zn},\label{eq:fluidb}\\
&&-i\tomega {\eta_n\over c^2}+
\left({d\over dr}\ln r\sigma +{n\mu\over r}\right)u_{rn}+
{\mu\over r}u_{r,n-2}+{d\over dr} u_{rn}\nonumber\\
&&\qquad\qquad\qquad +{im\over r}u_{\varphi n}
-{n\over h}u_{zn}=0,
\label{eq:fluid2}\ea
where 
$\mu\equiv  {d\ln h/d\ln r}$.
Eliminating $u_{\varphi n}$ and $u_{zn}$ from 
eqs.~(\ref{u1})-(\ref{eq:fluid2}),
we have
\ba
&&{d\eta_n\over dr}={2m\Omega\over r\tomega}\eta_n-{D\over\tomega }iu_{rn}
+{\mu\over r}[n\eta_n+(n+1)(n+2)\eta_{n+2}]\nonumber\\
&& \qquad +f_{rn}
+i{2\Omega\over\tomega}f_{\varphi n},\label{eq:dwndr}\\
&&{du_{rn}\over dr}=-\left[{d\ln(r\sigma)\over dr}+
{m\kappa^2\over 2r\Omega\tomega}\right]u_{rn}+{1\over i\tomega}
\left({m^2\over r^2}+{n\over h^2}\right)\eta_n\nonumber\\
&&\qquad +{i\tomega\over c^2}\eta_n
-{\mu\over r}(n u_{rn}+u_{r,n-2})
+{m\over \tomega r}f_{\varphi n}+{in\over\tomega h}f_{zn},
\label{eq:durndr}
\ea
where we have defined
\be
D\equiv \kappa^2-\tomega^2=\kappa^2-(\omega-m\Omega)^2.
\ee
We can check that for an external potential force, ${\bf f}=-\nabla
\phi$, these equations reduce to those of Zhang \& Lai (2006).

Consider local free wave solution of the form
$\eta_n\propto \exp\left[i\int^r\!k(s) ds\right]$.
For $|kr|\gg 1$, from the eqs.~(\ref{eq:dwndr})-(\ref{eq:durndr}),
in the absence of the external force, we find (see Okazaki et al.~1987; 
Kato 2001)
\be
(\tomega^2-\kappa^2)(\tomega^2-n\Omega_\perp^2)/\tomega^2=k^2c^2,
\label{eq:disp}\ee
where we have used $h=c/\Omega_\perp\ll r$ (thin disk),
and $m,n\ll r/h$. The dispersion relation is useful for identifying 
wave propagation zones in the disk (see Zhang \& Lai 2006).


\section{Waves at Lindblad/Vertical Resonance and Corotation Resonance}

We now consider the magnetic force on the disk as given in \S 2, i.e.,
the force per unit area is $F_z=F_\omega(r)\exp(i\varphi-i\omega t)$, 
with $\omega=\omega_s$ or $2\omega_s$. This corresponds to 
$m=n=1$, with
\be 
f_{z1}(r)={1\over\sigma}F_\omega(r),\quad f_r=f_\varphi=0,\qquad
(\omega=\omega_s,\,2\omega_s).
\ee
The wave propagation diagram [based on the dispersion relation 
(\ref{eq:disp}) with $n=m=1$] is shown in Fig.~1.

Note that for $\omega\ll \Omega=\kappa=\Omega_\perp$, the $m=n=1$
mode dispersion relation (\ref{eq:disp}) 
reduces to $\omega =\pm c k/2$, i.e., 
the bending wave propagates non-dispersively at speed $c/2$. This is the 
regime explored in a number of previous works (e.g., Papaloizou \& Lin 1995;
Ogilvie 2006). However, here we are particularly interested in the 
disk region where $\Omega$ is of the same order as $\omega$.

\begin{inlinefigure}
\vskip -2.cm
\hskip -2.cm
\scalebox{1.2}{\rotatebox{0}{\plotone{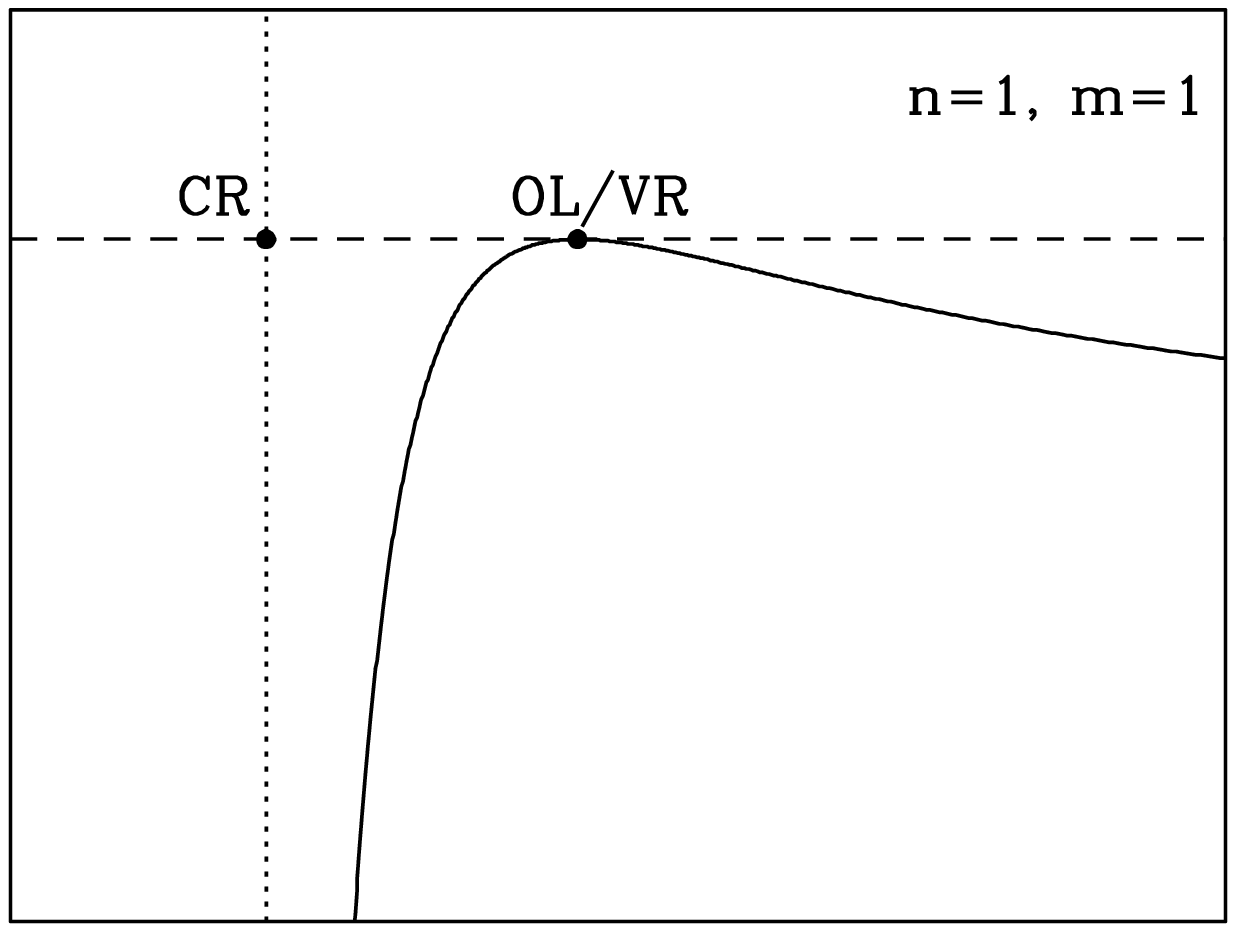}}}
\vskip -2.8cm
\figcaption{
A sketch of the function $G=(\kappa^2-\tomega^2)(1-\Omega_\perp^2/
\tomega^2)$ as a function of $r$ for $m=n=1$. The WKB dispersion relation is
$G=-k^2c^2$, and thus waves propagate in the region $G<0$. Note that 
the Lindblad/vertical resonance (L/VR) is the turning point $k\rightarrow 0$
and the corotation resonance (CR) is a singularity ($k\rightarrow\infty$).
}
\end{inlinefigure}

Consider eqs.~(\ref{eq:dwndr}-(\ref{eq:durndr}) for $n=1$. 
The coupling terms
can be neglected if $|\eta_1|\go |\eta_3|$, which is justified since
there is no $n=3$ driving force. Keeping only the potentially singular terms
($\propto D^{-1},\,\tomega^{-2}$), we have 
\ba
&& \Biggl[{d^2\over dr^2}+\left({d\over dr}\ln{r\sigma\over D}\right)
{d\over dr}
-{D(\tomega^2-n\Omega_\perp^2)\over h^2\Omega_\perp^2\tomega^2}\nonumber\\
&&\qquad\qquad -{2\Omega\over r\tomega}\left({d\over dr}\ln{\sigma\Omega
\over D}\right)\Biggr] \eta_1
={D\over \tilde{\omega}^2h}f_{z1}
\label{main1}\ea
This equation can be simplified further by changing variable via
\be
\eta_1(r)=\left({-D/r\sigma}\right)^{1/2}y(r).
\label{change}\ee
Keeping the leading-order terms, we obtain
\ba
&&\left[\frac{d^{2}}{dr^{2}}
-\frac{(\tilde{\omega}
^{2}-\Omega_{\perp }^{2})D}
{h^2\tilde{\omega}^{2}\Omega _{\perp }^{2}}
-\frac{3(\tomega d\Omega/dr+\kappa d\kappa/dr)^2}{D^2}
\right]y\nonumber\\
&&\qquad\qquad =-\frac{(-r\sigma D)^{1/2}}{h\tilde{\omega}^{2}}
f_{z1}.\label{y}\ea
The terms dropped inside $[...]$ in the above equation are 
considered to be smaller than the retained
terms for the following reasons: Firstly, away from
the $D=0$ region, they are negligible compared
to the term $\propto h^{-2}$ owing to the assumed smallness of $h$, 
and secondly, they are insignificant with respect
to the term $\propto D^{-2}$ in the vicinity of $D=0$.

Equation (\ref{y}) reveals two special points: one is at 
${\tilde{\omega} }=0$ (corotation resonance), 
the other is at $\tilde{\omega}^{2}=\kappa^{2}$ (Lindblad resonance).
The corotation resonance is a singularity of the wave equation, and 
the singularity implies that a steady emission or absorption of waves 
may occur there. The Lindblad resonances are turning points at which 
waves are generated or reflected.
Another interesting point is at 
$\ \tilde{\omega}^{2}=\Omega _{\perp }^{2}$.
It is a transition point where the vertical resonance occurs.
For a  Keplerian disk, $ \kappa =\Omega _{\perp}=\Omega $, the 
vertical resonance coincides with the
Lindblad resonance, and we will call it Lindblad/vertical resonance
(L/VR). Equation (\ref{y}) then reduces to
\ba
&&\left[\frac{d^{2}}{dr^{2}}
+\frac{(\tilde{\omega}^{2}-\Omega^{2})^2}
{h^2\tilde{\omega}^{2}\Omega _{\perp }^{2}}
-\frac{3\omega ^{2}(d\Omega/dr)^{2}}{(\tilde{\omega}^{2}-\kappa^{2})^{2}}
\right]y\nonumber\\
&&\qquad =-\frac{\sigma^{\frac{1}{2}}r^{\frac{1}{2}}
(\tilde{\omega}^{2}-\kappa ^{2})^{1/2}}{h\tilde{\omega}^{2}}
f_{z1}\equiv S_\omega(r).
\label{w}\ea
This is our basic working equation.

\subsection{Lindblad/Vertical Resonance}

We now study wave excitations near a Lindblad/vertical resonance
(L/VR), where $D=0$,
or $\tomega=\kappa=\Omega$, $\omega=2\Omega$, and the (outer) L/VR
radius is denoted by $r_L$. Changing $r$ to the new
variable $x\equiv (r-r_{L})/r_{L}$ and keeping the leading order terms,
we find that, for $|x|\ll 1$, eq.~(\ref{w}) reduces to
\be
\frac{d^{2}}{dx^{2}}y+\left(b^2x^{2}-\frac{3}{4x^2}\right)y=
S_\omega r_L^2,\label{yl}\ee
where 
\be
b^2=\frac{64}{\omega^{2}}\frac{(d\Omega /dr)^{2}r^{4}}{h^{2}}
\Biggl |_L=36\left({r\over h}\right)^2_L
\ee
(The subscript ``L'' implies that the quantity should be evaluated
at $r=r_L$). The two independent solutions of the homogeneous version of 
eq.~(\ref{yl}) are
\be 
y_{\pm}\propto (bx)^{-1/2}\,e^{\pm ibx^2/2}.
\label{eq:yll}\ee
The general solution of the inhomogeneous eq.~(\ref{yl}) is then
\be
y=-y_+\int^x {y_-S_\omega r_L^2\over W}\,dx
+y_-\int^x {y_+S_\omega r_L^2\over W}dx
+C_+ y_+ +C_- y_-,
\ee
where $W=y_+dy_-/dx-y_-dy_+/dx=-2i$ is the Wronskian. The constants 
$C_{\pm}$ can be fixed by requiring waves propagating away from the
resonance. Using eq.~(\ref{change}), we obtain
\ba
&&\eta_1=i\left({rh\over 24}\right)^{1/2}_Lf_{z1}
\Biggl[e^{i\zeta^2/2}\int^\zeta_{-\infty} e^{-i\zeta^2/2}d\zeta
\nonumber\\
&&\qquad +e^{-i\zeta^2/2}\int^{\infty}_\zeta e^{i\zeta^2/2}d\zeta\Biggr],
\label{ynh}\ea
where $f_{z1}$ is evaluated at $r=r_L$, and we have defined $\zeta=b^{1/2}x$.
Note that although our analysis here is limited to $|x|\ll 1$, we have 
extended the integration limit to $\zeta=\pm\infty$ in eq.~(\ref{ynh}).
This is valid because $b=6r_L/h\gg1$ for a thin disk and 
the integrands in the integrals are highly oscillatory for $|\zeta|\gg 1$
(so that the contribution to the integrals from the $|\zeta|\gg 1$ region 
is negligible).

To calculate the angular momentum transfer through the resonance,
we note that for $\zeta\rightarrow +\infty$ (but still $x\ll 1$),
equation (\ref{ynh}) becomes
\be
\eta_1=i\left({rh\over 24}\right)^{1/2}_L\!f_{z1}\,
(-2\pi i)^{1/2}\,e^{i\zeta^2/2}\qquad (\eta\rightarrow\infty).
\label{eq:pinf}\ee
The angular momentum flux to the $r>r_L$ is 
\ba
&&F(r>r_L)={\pi r\sigma\over D}{\rm Im}\left(\eta_1{d\eta_1^\ast\over dr}
\right)\nonumber\\
&&\quad=-{\pi^2\over 2}\left({\sigma\,r\over dD/dr}|f_{z1}|^2\right)_L
={\pi^2\over 3\omega^2}\left(r^2\sigma |f_{z1}|^2\right)_L.
\ea
Similarly, for $\zeta\rightarrow -\infty$, equation (\ref{ynh}) gives
\be
\eta_1=i\left({rh\over 24}\right)^{1/2}_L\!f_{z1}\,
(2\pi i)^{1/2}\,e^{-i\zeta^2/2}\qquad (\eta\rightarrow -\infty).
\label{eq:minf}\ee
The angular momentum flux to the $r<r_L$ region is identical to
$F(r>r_L)$. The total torque on the disk acted through the OL/VR is then 
\be
T_L=2F(r>r_L)=
{2\pi^2\over 3\omega^2}\left(r^2\sigma |f_{z1}|^2\right)_L.
\label{eq:tl}
\ee

To estimate the torque $T_L$, we consider $\omega=\omega_s$, and
use $f_{z1}=F_\omega/\sigma$ with $F_\omega\sim \mu^2/(4\pi r^6)$
(see \S 2).  For a Keplerian disk, $r_L=2^{2/3}r_c$ and
we assume $r_c/r_m=\lambda$ (for accretion to occur, we require $\lambda>1$).
Using eq.~(1), we then obtain
\be
T_L\sim 4\times 10^{-4}{\dot M^2\over (\xi\lambda)^7\sigma(r_L)}.
\ee
The canonical accretion torque is $T_A\equiv \dot M (GMr_m)^{1/2}$.
Using $\dot M=2\pi (r\sigma |u_r|)_L$ (where $u_r$ is the radial 
velocity if the unperturbed accretion flow), we find
\be
{T_L\over T_A}\sim 3\times 10^{-3}\xi^{-7}\lambda^{-6.5}
\left({|u_r|\over r\Omega}
\right)_L.
\ee
Thus the resonant torque is much smaller than the accretion torque.

\subsection{Corotation Resonance}

The corotation resonance (CR) is located where $\tomega=0$ or $\omega=
m\Omega(r_c)$. The WKB dispersion relation
shows that for $n=0$, waves are evanescent in the region around 
the corotation radius $r_c$, while for $n>0$ wave propagation is 
possible around $r_c$. 

We again focus on the $n=m=1$ waves.
In the vicinity of corotation,
the terms $\propto h^{-2}$ in eq.~(\ref{main1}) or (\ref{w}) 
are dominant, and we only need to keep these terms and the 
second-order differential term.
Equation (\ref{main1}) then reduces to 
\be
{d^2\over dr^2}\eta_1
-{D(\tomega^2-\Omega^2)\over h^2\tomega^2\Omega_\perp^2}\eta_1
={D\over h\tomega^2}f_{z1}.
\label{eq:a3}\ee
Defining $x=(r-r_c)/r_c$ and expanding eq.~(\ref{eq:a3}) around $x=0$, we have
\be
\left[{d^2\over dx^2}+{C\over (x+i\epsilon)^2}\right]\eta_1 
= C{hf_{z1}\over (x+i\epsilon)^2}
\label{c1}\ee
where $hf_{z1}$ is evaluated at $r=r_c$, and
\be
C=\left({\Omega\over h\,d\Omega/dr}\right)^2_{c}=
\left({2r\over 3h}\right)_{c}^2\gg 1
\ee
(where the subscript ``c'' means that the quantity is evaluated at $r=r_c$).
In eq.~(\ref{c1}), we have inserted a small imaginary part $i\epsilon$
(with $\epsilon>0$) in $1/x^2$ because we consider the response of the
disk to a slowly increasing perturbation.

The general solution to eq.~(\ref{c1}) is
\ba 
\eta_1 &=&hf_{z1}+Mz^{1/2}z^{i\nu}+Nz^{1/2}z^{-i\nu}\nonumber\\
&=& h f_{z1} +Mz^{1/2}e^{i\nu \ln z}+Nz^{1/2}e^{-i\nu \ln z},
\label{c10}\ea
where $\nu=\sqrt{C-{1\over4}}\gg 1$, $z=x+i\epsilon$ (with $\epsilon>0$)
and $M$ and $N$ are constants. The first term, the non-wave part, is a particular solution, 
while the other two terms are solutions to the homogeneous equation, 
depicting the waves. The $z^{1/2}z^{i\nu}$ term has a local wave number 
$k=d(\nu\ln z)/dr=\nu/(r_c x)$, with the group velocity
$v_g=d\omega/dk=-\tomega/k=-3r_cx^2\Omega_c/(2\nu)<0$, 
thus it represents waves propagating toward small $r$. Similarly, 
the $z^{1/2}z^{i\nu}$ term has $v_g>0$ and represents waves propagating 
toward large $r$. 

As shown in Zhang \& Lai (2006), waves with $n=1$ can propagate into the corotation 
region and get absorbed there. Consider 
an incident wave propagating from the $x>0$ region toward $x=0$:
\be 
\eta_1(x>0)=A_+\, x^{1/2}e^{i\nu \ln x},
\label{eq:incident}\ee
where $A_+$ is constant specifying the wave amplitude.
The transmitted wave is given by
\be
\eta_1(x<0)=i\, A_+\, e^{-\pi\nu} (-x)^{1/2}e^{i\nu\ln (-x)}.
\ee
Since $\nu\gg 1$, the wave amplitude is vastly decreased by a 
factor $e^{-\pi\nu}$ after propagating through the corotation. 
The net angular momentum flux absorbed at corotation is 
\ba
\Delta F_c=
\pi \left({\sigma\over \kappa^2}\right)_{c}\!\nu |A_+|^2(1+
e^{-2\pi\nu})\simeq 
\left({2\pi r\sigma\over 3 h \omega^2}\right)_{c}|A_+|^2,
\label{deltafc}\ea
where in the last equality we have used $\nu\simeq \sqrt{C}\gg 1$.
Thus, a wave propagating from $r>r_c$ into the corotation resonance 
deposits almost all of its positive angular momentum at $r=r_c$.
Similarly, a wave propagating from $r<r_c$ into the corotation resonance 
deposits its negative angular momentum at $r=r_c$.

\section{Global Disk Response}

Having studied the behavior of waves near L/VR and CR, we can now
construct global solution for the waves excited by the external
force.

Away from the L/VR, the two linearly independent WKB solutions to the 
homogeneous wave eq.~(\ref{w}) are
\be
y_\pm=k_r^{-1/2}\exp\left(\pm \,i\int^r k_r\,dr\right),
\label{eq:ypm}\ee
where 
\be
k_r\equiv {\tomega^2-\Omega^2\over h\Omega\tomega}
={\omega(\omega-2\Omega)\over h\Omega(\omega-\Omega)}.
\ee
It is easy to check that eq.~(\ref{eq:ypm}) reduces to 
(\ref{eq:yll}) for $|r-r_L|\ll r_L$, 
thus it is valid even near the L/VR. From eq.~(\ref{change}), we have 
\be
\eta_{1\pm}=\left[{h\Omega(\omega-\Omega)\over r\sigma}\right]^{1/2}
\exp\left(\pm \,i\int^r k_r\,dr\right).
\label{eq:etapm}\ee
Near the CR, this solution reduces to
\be
\eta_{1\pm}\propto x^{1/2}e^{\mp i\nu\ln x}, \qquad {\rm for}~~
|x|=|(r-r_c)/r_c|\ll 1.
\ee
Thus equation (\ref{eq:etapm}) represents the two independent solutions of the
homogeneous wave equation for all radii. 

As discussed before, Waves are mainly excited at the L/VR, where 
waveform is given by eq.~(\ref{ynh}). Matching eq.~(\ref{eq:pinf}) 
or (\ref{eq:minf}) 
with the general solution, $\eta_1=C_+\eta_{1+}+C_-\eta_{1-}$ 
(where $C_\pm$ are constants), we find that the waves away from the
L/VR are:
\ba
&&\eta_1=\left({i\pi r^2\sigma\over 3\omega^2}\right)^{1/2}_L\!f_{z1}(r_L)\,
\left[{h\Omega(\omega-\Omega)\over r\sigma}\right]^{1/2}\nonumber\\
&&\qquad\qquad \times\exp\left(\,i\int^r_{r_L} k_r\,dr\right),\qquad (r>r_L)
\label{eq:eta111}\\
&&\eta_1=\left({-i\pi r^2\sigma\over 3\omega^2}\right)^{1/2}_L\!f_{z1}(r_L)\,
\left[{h\Omega(\omega-\Omega)\over r\sigma}\right]^{1/2}\nonumber\\
&&\qquad\qquad \times
\exp\left(\,-i\int^r_{r_L} k_r\,dr\right),\quad (r_c<r<r_L).
\label{eq:etarcrl}
\ea
Note that although for $r_c<r<r_L$ the wave (\ref{eq:etarcrl})
has positive phase velocity (since $k_r<0$ for $r_c<r<r_L$),
the group velocity is negative, i.e. the wave propagates toward small 
radii. As this wave approaches $r_c$, the waveform reduces to
[see eq.~(\ref{eq:incident})]
\be
\eta_1=A_+x^{1/2}\,e^{i\nu\ln x},
\qquad
\left(0<x={r-r_c\over r_c}\ll 1\right)
\ee
where 
$\nu={2r_c/(3h_c)}$
and 
\ba
&& A_+=\left(-i\pi r^2\sigma\right)^{1/2}_L\!f_{z1}(r_L)\,
\left({h\over 2r\sigma}\right)_c^{1/2}\nonumber\\
&&~~\qquad \times\exp\left(-i\int^{r_+}_{r_L} k_r\,dr
-i\,\nu\,\ln x_{+}\right),
\ea
where $0<x_+\equiv (r_+-r_c)/r_c\ll 1$.
After passing through the CR, the wave amplitude is significantly reduced:
\ba
&&\eta_1=i\, A_+\, 
e^{-\pi\nu}(-x)^{1/2}\,
e^{i\,\nu\ln (-x)},\nonumber\\
&&\qquad\qquad\quad
\left(x={r-r_c\over r_c}<0,~~~|x|\ll 1\right).
\ea
This then joins onto the inward-going wave solution
\ba
&&\eta_1=e^{-\pi\nu}\,
\left({-i\pi r^2\sigma\over 3\omega^2}\right)^{1/2}_L\!f_{z1}(r_L)
\left[{h\Omega(\omega-\Omega)\over r\sigma}\right]^{1/2}\nonumber\\
&&\qquad\times \exp\left(-i\int^r_{r_-} k_r\,dr-i\varphi\right),
\qquad (r<r_c),
\ea
with $\varphi=\int^{r_+}_{r_L} k_r\,dr
+\nu\ln (-x_{+}/x_-)$, where $x_-=(r_--r_c)/r_c<0$ and $|x_-|\ll 1$.
Using eq.~(\ref{deltafc}), we find that the angular momentum flux 
deposited at $r_c$ is 
\be
\Delta F_c={\pi^2\over 3\omega^2}\left(r^2\sigma |f_{z1}|^2\right)_L,
\ee
in agreement with the angular momentum flux 
emitted from $r_L$ toward corotation
[see eq.~(\ref{eq:tl})].

Figure 2 depicts the global solution describing waves excited at 
the L/VR, which either propagates outwards or inwards toward the CR
at which the angular momentum deposition occurs.

\begin{inlinefigure}
\scalebox{1.1}{\rotatebox{0}{\plotone{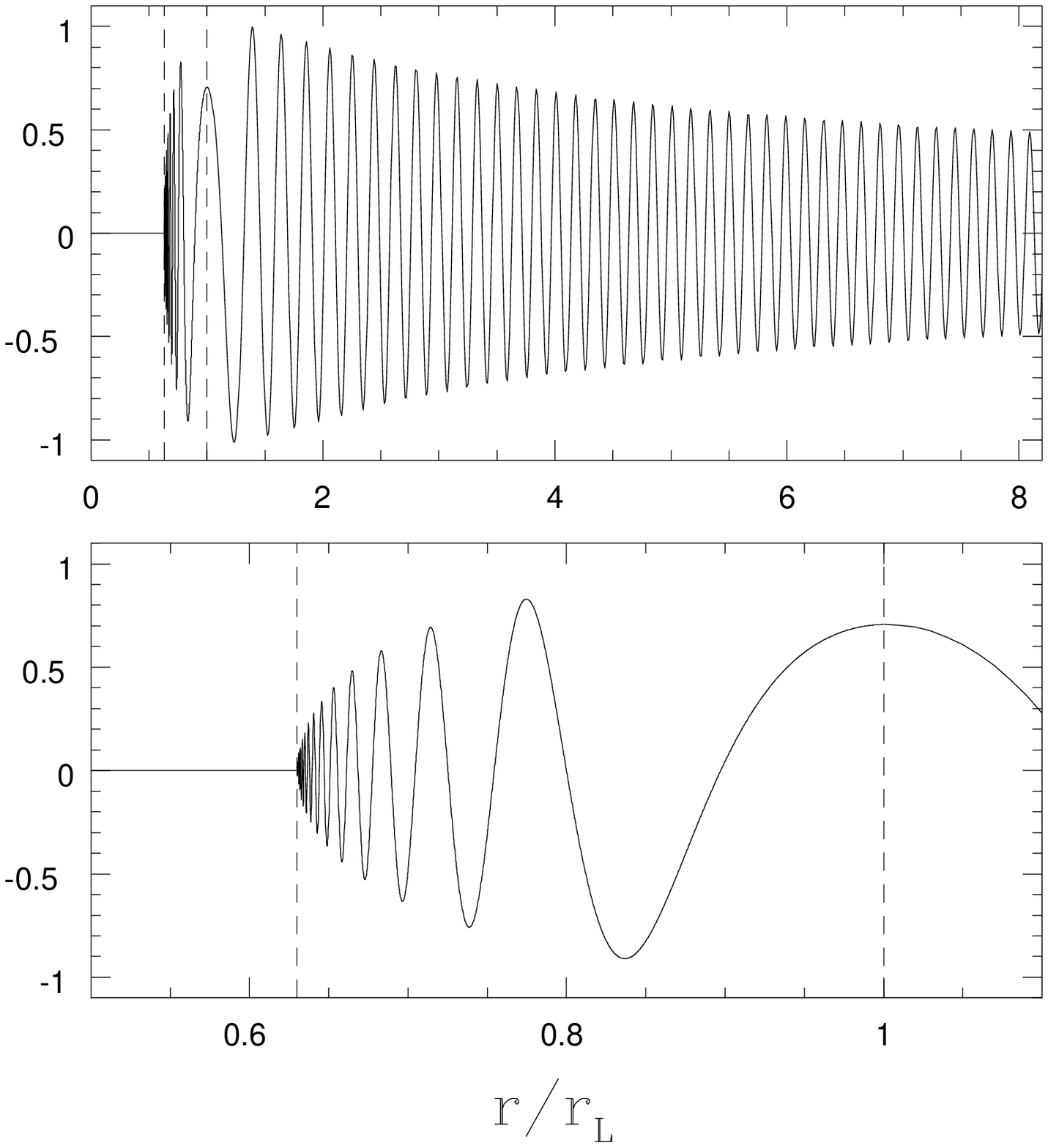}}}
\figcaption{
The enthalpy perturbation $\eta_1$ of the wave driven by 
an external force ($m=n=1$). The wave is excited 
at the L/VR ($r/r_L=1$), and propagates toward large radii and small radii. 
The inward-going wave is absorbed at the corotation resonance. The lower 
panel focuses on the region between corotation and $r_L$.
}\end{inlinefigure}

\section{Application to QPOs}

The main result of the paper is illustrated in Fig.~2. It shows that 
under a periodic magnetic vertical forcing, the disk responds by launching 
$m=n=1$ waves from the Lindblad/vertical resonance (where $\omega=2\Omega$,
with the forcing frequency $\omega=\omega_s$ or $2\omega_s$).
The wave either propagates to large radii or propagates inward
where it gets absorbed at corotation resonance (where $\omega=\Omega$).

We have already shown that the torque carried by the excited
wave is small compared to the canonical accretion torque (see \S 4.1).
Nevertheless the wave may manifest itself by inducing or modulating
variabilities.
Clearly, the wave is most visible at the Lindblad/vertical resonance
since the wavelength is the largest there. Variations caused by the wave
are most likely due to fluid elements around $r_L$. From 
eqs.~(\ref{eq:eta111})-(\ref{eq:etarcrl}), we find that the amplitude of
enthalpy perturbation at $r_L$ is given by
$|\eta_1|=(\pi r/12 h)^{1/2}(F_\omega h/\sigma)$ (all quantities are evaluated
at $r=r_L$). With $F_\omega\sim \mu^2/(4\pi r^6)$ (see \S 2), 
$r_m=r_c/\lambda$ (where $\lambda>1$ is a constant),
 $r_L=2^{2/3}r_c$, and using eq.~(1) and
$\dot M=2\pi r\sigma |u_r|$,  we find 
the dimensionless amplitude
\be
{|\eta_1|\over c^2}\sim 0.05\,(\lambda\xi)^{-7/2}\left({r u_r^2\over h c^2}
\right)^{1/2}_L,
\ee
where $u_r$ is the radial (inflow) velocity. For a $\alpha$-disk, 
$|u_r|\simeq (3/2r)\alpha hc$ not too close to the disk inner edge, we have
\be
{|\eta_1|\over c^2}\sim 0.1\,\alpha\,(\lambda\xi)^{-7/2}\left({h\over r}
\right)^{1/2}_L.
\ee
The dimensionless vertical velocity perturbation, $|u_{z1}|/c$, 
is the of the same order of magnitude 
at $r\sim r_L$ as $|\eta_1|/c^2$.
Since $\lambda\xi\sim 1$, the dimensionless perturbation amplitude
may reach a few percent, while definitely remaining in the linear regime.
Note that the vertical velocity perturbation amplitude increases
toward the corotation radius, non-linear saturation will occur there.
But since the wavelength approaches
zero as $r\rightarrow r_c$, variations caused by fluid elements around 
corotation resonance may not be visible.

It seems inevitable that bending waves studied in this paper
will be excited in disks around magnetic stars.
We may suggest various ways that these resonantly excited waves may 
induce variabilities and QPOs. In the context of LMXBs
(van der Klis 2006), one may 
imagine that the oscillating fluid perturbation at the Lindblad resonance
can produce a beat phenomenon by modulating the ``seed'' radiation from the 
inner region of the disk (e.g. Lamb \& Miller 2003). 
Suppose the ``seed'' radiation has a quasi-periodic
variability with frequency $\nu_h$ (e.g. due to orbital motion of blobs
at the disk inner edge or some other mechanisms). The fluid element
at the Lindblad/vertical resonance has orbital frequency $\Omega=\omega/2$
(since $\omega-\Omega=\Omega_\perp$ for $m=1$, and $\Omega_\perp\simeq
\Omega$). The reprocessed radiation would then show an additional QPO
at frequency $\nu_l=\nu_h-\nu/2$.
In this regard, it is particularly interesting that in our model, the 
driving frequency $\nu=\omega/(2\pi)$ can be either the spin frequency
$\nu_s$ or twice of that, depending on the disk geometry with respect to
the stellar spin and dipole axes (see \S 2). As noted in \S 1,
both $\nu_h-\nu_l=\nu_s$ and $\nu_h-\nu_l=\nu_s/2$ have been
observed in LMXBs. This feature has not been explained by current models
(see van der Klis 2006). For many systems, the beat is not perfect. This might be
due to the fact the excited wave still has significant strength
away from the Lindblad/vertical resonance (see Fig.~2). 

Finally we note that our treatmnet of wave excitation by magnetic
forces 
is performed under simplifying assumptions, as detailed in section 3. 
We hope that the novel features of our model in connection with QPOs
will motivate future, more rigorous studies of the dynamics of waves
in disks around magnetic stars.

\acknowledgments
This work has been supported in part by NASA Grant NNX07AG81G and by NSF
grant AST 0707628.

\end{document}